\documentstyle[12pt]{article}

\newcommand{\be}{\begin{equation}}
\newcommand{\ee}{\end{equation}}
\newcommand{\ba}{\begin{eqnarray}}
\newcommand{\ea}{\end{eqnarray}}
\newcommand{\pa}{\partial}

\begin{document}
\begin{titlepage}
\begin{flushright}
{IFUM 471/FT}\\
{ROM2F-94/43}\\
\hfill \\
{October \ 1994}
\end{flushright}
\vskip 8mm
\begin{center}

{\large \bf    Shock Waves and the Vacuum Structure of Gauge
Theories\footnote{Contribution to ``Quark Confinement and
the Hadron Spectrum'', Como 20-24 June 1994.}}\\

\vspace{1cm}

{\bf Maurizio \ Martellini}

\vspace{0.3cm}

{\sl Dipartimento di Fisica\\
Universit\`a di Milano\\
Via Celoria 16 \ \ 20133 \ Milano \ \ ITALY \\
I.N.F.N.\ - \ Sezione di Pavia }

\vspace{.6cm}

{\bf Augusto Sagnotti}
\vspace{0.3cm}

{\sl Dipartimento di Fisica\\
Universit\`a di Roma \ ``Tor Vergata'' \\
I.N.F.N.\ - \ Sezione di Roma ``Tor Vergata'' \\
Via della Ricerca Scientifica, 1 \ \
00133 \ Roma \ \ ITALY}

\vspace{.6cm}

{\bf Mauro Zeni}

\vspace{0.3cm}

{\sl Dipartimento di Fisica \\
Universit\`a di Milano \\
I.N.F.N. \ - \  Sezione di Milano \\
Via Celoria 16 \ \ 20133 \ Milano \ \ ITALY}

\vspace{.6cm}
\end{center}

\abstract{In Yang-Mills theory massless point sources lead naturally to
shock-wave configurations.  Their magnetic
counterparts endow the vacuum of the four-dimensional compact abelian
model with a Coulomb-gas behaviour whose physical implications are briefly
discussed.
}
\vfill
\end{titlepage}
\addtolength{\baselineskip}{0.3\baselineskip}

The current semiclassical picture of the vacuum in gauge
theories rests, to a large extent, on the known
solutions of the Yang-Mills field equations\cite{actor}. It is common wisdom
that attaining a detailed understanding of this vacuum is a major challenge for
Quantum Field
Theory, as well as a crucial step in assessing its actual role in
High-Energy Physics. In this talk we consider a class of rather
simple shock-wave solutions of the field
equations with massless point sources. In the compact abelian gauge theory,
their euclidean counterparts exhibit rather neatly a phase
transition\cite{kogutlh}, thus providing a simple explicit realization of the
standard picture\cite{thooft}.

Let us begin by considering the Yang-Mills field equation $D^{\mu}
F_{\mu\nu}=4\pi
j_{\nu}$ with the massless point source
\be
   j_{\nu}^a \ = \ q \ I^a \ \delta_ {\nu}^u \
   \delta (u) \ \delta^2 ( {\bf r}) \qquad , \label{jel}
\ee
where $u$ and $v$ are
light-cone coordinates ($u={x^0-x^3\over \sqrt{2} }$ ;  $v={x^0+x^3\over
\sqrt{2}}$), ${\bf r}$ is a space-like coordinate
vector orthogonal to $u$ and $v$, and $I^a$ labels
the color charge of the point particle. The resulting classical solution,
\be
   {A^a}_{\mu} \ = \ - \ 2 \ q \ I^a \ \delta^u_{\mu} \ \delta (u)
   \ \log r \quad ,\label{ael}
\ee
where $r$ denotes the length of ${\bf r}$,
may be obtained by inspection, or by a simple extension of the ``cut and
paste''
procedure\cite{dray} used to generate a similar type of gravitational wave.
The electric and magnetic field strengths associated to eq. (\ref{ael}),
\ba
   E_i &\! =&\! \ {\sqrt{2} \ q} \ I^a \ \delta (u) \ {r^i\over r^2}
\qquad , \\ \nonumber
   B_i &\! =&\! - \ {\sqrt{2} \ q} \ I^a \ \delta (u) \ \varepsilon^{ij} \ {
r^j\over r^2} \qquad\qquad
   (i = 1,2 ) \qquad , \label{ebel}
\ea
may also be obtained as singular boosts of a static Coulomb field in the limit
where the velocity $v/c \ \rightarrow 1$  \cite{jackiw}. The
``two-dimensional''
shape of the shock wave applies to arbitrary
superpositions of comoving massless point-like currents and may be ascribed to
the relativistic contraction of the fields in the longitudinal direction.

In the quantum theory, the topological character
of the field configuration space plays a crucial
role in determining the nature of the correct degrees of
freedom.  In the prototype example for this type of phenomenon, the 2D XY
model, the fundamental field is an angular variable $\theta({\bf r})$ and the
naive elementary excitations, long-wavelength spin waves, must be supplemented
with genuinely new ones, vortices. From the mathematical viewpoint, vortices
are
singular field configurations that result in effective (quantized) violations
of the ``Bianchi identity'' for $d \theta$.
These excitations are responsible for the Kosterlitz-Thouless ($KT$) phase
transition\cite{kt} that separates the weak and strong coupling regimes of the
model.

Vortices are actually relevant in a number of different models with periodic
field configuration spaces, most notably in lattice gauge theories,
where they may be associated with
magnetic monopoles. Again, in continuum formulations they manifest
themselves as singular field
configurations that violate the relevant ``Bianchi identities''. A number of
reasons call for a periodic formulation of abelian gauge theories, most notably
the fact that in unified models abelian gauge fields typically emerge from the
spontaneous breaking of non-abelian gauge symmetries\cite{thooft}.

For
compact 3D QED Polyakov\cite{polyakov} has shown that the
periodicity results in a vacuum filled with a plasma of monopoles and in a mass
gap for the gauge fields, as well as in a confinement phenomenon at all scales.
For compact 4D QED a quantitative discussion of the role of vortices
has been
hampered by the string-like nature of the corresponding excitations,
monopole current loops. Still, the basic argument is rather simple. It is
best exhibited starting
from the Villain\cite{villain} form of the action
\be
   S={1\over 4g^2}
   \sum_{x,\alpha\beta} (F_{\alpha\beta}(x) +2\pi n_{\alpha\beta}(x) )^2
   \quad ,\label{svil}
\ee
where the gauge potential is an
angular variable and the
$n_{\alpha\beta}(x)$ are matrices of integers associated to the lattice
plaquettes that may be related to integer-valued currents upon
integration over elementary cubes  $\Box^{\mu}$ of the lattice,
\be
   \oint_{\pa\Box^{\mu}} n_{\alpha\beta}(x) = j^{\mu}\quad . \label{cha}
\ee

The Berezinsky decomposition
$n(x)_{\alpha\beta}
   =\pa_{\alpha} n_{\beta} \ - \ \pa_{\beta} n_{\alpha} \ + \
\varepsilon_{\alpha\beta\gamma\delta} \ \pa_{\gamma} \varphi_{\delta}$
exhibits the longitudinal part of $n$, as well as the
vector field $\varphi_{\mu}$, whose source is the current of eq.
(\ref{cha}) and whose (dual) field strength adds to the usual one. Since
the total field strength
\be
   F^{\prime}_{\alpha\beta} \ = \ F_{\alpha\beta} \ + \ 2\pi
n_{\alpha\beta} \ = \ F_{\alpha\beta} \ + ... + \ {1 \over 2} \
\varepsilon_{\alpha\beta\gamma\delta}
 \ {\cal F}_{\gamma\delta} \label{bia}
\ee
violates the Bianchi identity of the original gauge field, the current in
eq. (\ref{cha}) is of {\it magnetic} origin.

An elegant lattice description of
this model, allowing for electric current (Wilson) loops as well, was presented
in \cite{bmk}. The resulting picture involves
interacting  monopole and charge loops, but it
is difficult to turn it into a quantitative analysis, since in this case the
effective sine-Gordon description of 3D QED\cite{polyakov} should be replaced
by a theory of monopole loops.  Although a proper description of
this theory is likely to be complicated, the
intuition gained from ordinary string theory\cite{gsw} suggests a possible way
of gaining quantitative insight
into the problem. This may be associated to ``straight'' current loops, and
conceivably to massless ones near the phase transition, that according to
numerical estimates appears to be of second order\cite{kogutlh}.

In the remainder of this talk we would like to show how the truncation
of the monopole strings to these ``zero modes'' accounts both for the phase
transition and for Wilson's area law\cite{wilson} in a rather neat fashion. To
this end, we need the duals of the
fields in eq. (3). They may be derived from a ``magnetic'' analogue
of the vector potential of eq. (\ref{ael}) or, alternatively, from the ordinary
potential
 \be
   A_{\mu} = (0,0,A_i ) \ ,\qquad {\rm where} \qquad
   A_i \ = \ q_{(m)} \theta (u) \ \varepsilon^{ij} \
   {r^j\over r^2} \qquad ( i,j= 1,2 ) \quad , \label{ama}
\ee
whose Dirac string
\be
   B_3 \ = \ 2 \pi q_{(m)} \theta (u) \ \delta^2 ({\bf r}) \quad  \label{dstr}
\ee
would be ineffective in eq. (\ref{svil}) upon suitable quantization of the
monopole charge,
a single Dirac quantum corresponding to $q_{(m)} \pm 1$.
Interestingly, the potential in eq. (\ref{ama}) plays an important role in knot
theory\cite{khono}.

In computing the action for a pair of waves, it is
convenient to resort to a covariant notation, that has the
further advantage of illuminating the geometry of the space-like planes where
the interactions take place.  A ``straight'' current in a generic
direction may be written
\be
   j_{\mu}(x) \ = {1 \over 2} \ q_{(m)} k_{\mu} \int d\tau
   \delta^4 [x^{\alpha}-x_0^{\alpha}-k^{\alpha}\tau ] \quad ,\label{jcov}
\ee
where in the massless case $k_{\mu}$ is a null vector. The total action
for a pair $(ij)$ of shock waves is then
\be
   S_{ij} \ = \ {\pi q_{(m)i}q_{(m)j}\over g^2} \
   \log ( x_{ij} \ \Pi \ x_{ij} ) \quad , \label{loga}
\ee
where $x_{ij}$ is the distance between the two lines and
$\Pi$ is a projector onto the space-like plane orthogonal to the two
wave vectors. This result essentially holds in the Wick rotated
case as well, where the calculation requires a suitable extension of
$\delta(x)$ to the complex plane, rather interesting in its own right.
The logarithmic interaction closely parallels the
state of affairs for the XY model and is just enough to
compete with the point-like entropy of these configurations, thus displaying a
$KT$-like phase transition, while the
divergent self interactions require globally neutral
sets of monopole currents. Interestingly, a transition of this type
would be predicted by the Migdal-Kadanoff approximation\cite{mk}, known to
become less accurate as the space dimensionality is increased, consistenly with
our neglect of higher extended excitations. The naive estimate of the
transition temperature from eq. (\ref{loga}) is rather amusing, since it yields
$2{g^2}_c = \pi/2$ for a pair of fundamental Dirac monopoles with a $KT$-like
measure $d^4 x$, to be
compared with the loop-space estimate for the full gas of monopole
strings\cite{bmk}\cite{kogutlh}, $2 {g^2}_c
\approx 1.57$!

Finally, the area law\cite{wilson} may be
anticipated by comparing the effective KT structure of our
``straight-line'' vacuum to the Coulomb-gas picture of the
three-dimensional model of ref.\cite{polyakov}.  One may then
arrive at an effective Sine-Gordon dynamics to infer that, above
the critical temperature $T_c$, double
layers of monopole lines form around the Wilson loop, thus implying the
area law and confinement.
One may also envisage a similar analysis of black-hole
physics\cite{dray}, given the formal similarity between the logarithm of
the Wilson loop and the Bekenstein entropy. More details, including
the explicit invariant measure for the
moduli, will be presented elsewhere\cite{msz}.

\vskip 24pt
\begin{flushleft}
{\Large \bf Acknowledgements}
\end{flushleft}

It is a pleasure to acknowledge stimulating discussions with T. Banks
and A. Di Giacomo. We also benefitted from the kind hospitality of the
Physics Departments of the University of Milan and of the University of Rome
``Tor Vergata'' and of the Centre de Physique Theorique of the
Ecole Polytechnique. This work was supported
in part by E.E.C. Grants CHRX-CT93-0340 and CHRX-CT92-0035.

\vfill\eject


\begin{thebibliography}{99}
\bibitem{actor} For a review, see A. Actor, {\it Rev. Mod. Phys.} {\bf 51}
(1979) 461.
\bibitem{kogutlh} For a review, see J.B. Kogut, in {\it Recent Advances in
Field Theory and Statistical Mechanics}, Proc. Les Houches 1982, eds.
J.B. Zuber and R. Stora (North-Holland, Amsterdam 1984).
\bibitem{thooft} G. `t Hooft, {\it Proc. 1975 EPS
Conference}, ed. A. Zichichi (Compositori, Bologna, 1976); \hfil\break
S. Mandelstam, {\it Phys. Rep.} {\bf 23C} (1976) 245;\hfil\break
A.M. Polyakov, {\it Gauge Fields and Strings} (Harwood, New
York, 1987).
\bibitem{dray} T. Dray and G. `t Hooft,
  {\it Nucl. Phys.} {\bf B253} (1985) 173.
\bibitem{jackiw} R. Jackiw, D. Kabat and M. Ortiz,
  {\it Phys. Lett.} {\bf B277} (1992) 148.
\bibitem{kt} J.M. Kosterlitz and D.J. Thouless, {\it J. Phys.} {\bf C6} (1973)
1181; \hfil\break J.M. Kosterlitz, {\it J. Phys.} {\bf C7} (1974) 1046.
\bibitem{polyakov} A.M. Polyakov, {\it Phys. Lett.} {\bf B59} (1975) 79.
\bibitem{villain} J. Villain, {\it J. Physique} {\bf 36} (1975) 581.
\bibitem{gsw} M.B. Green, J.H. Schwarz and E. Witten, {\it Superstring
Theory} (Cambridge University Press, Cambridge, 1987).
\bibitem{khono} T. Khono, {\it Adv. Studies in Pure Math.} {\bf 16}
(1988) 255;\hfil\break
E. Guadagnini, M. Martellini e M. Mintchev,
{\it Nucl. Phys.} {\bf B336} (1990) 581.
\bibitem{wilson} K.G. Wilson, {\it Phys. Rev.} {\bf D10} (1974) 2445.
\bibitem{bmk} T. Banks, R. Myerson  and J. Kogut,
{\it Nucl. Phys.} {\bf B129} (1977) 493; \hfil \break
A. Ukawa, P. Windey and A.H. Guth, {\it Phys. Rev.} {\bf D21}
(1980) 1013.
\bibitem{mk}A.A. Migdal, {\it Sov. Phys. JETP} {\bf 42} (1975) 413;\hfil \break
L.P. Kadanoff, {\it Rev. Mod. Phys.} {\bf 49} (1977) 267.
\bibitem{msz} M. Martellini, A. Sagnotti and M. Zeni, in preparation.
\end{thebibliography}
\end{document}